\documentclass[fdp,fleqn]{w-art}
\usepackage{times}
\usepackage{w-thm}
\usepackage[]{graphicx}
\usepackage[]{amssymb}
\usepackage{epic}
\usepackage[]{mathrsfs}
\begin{document}
\DOIsuffix{theDOIsuffix}
\Volume{x}
\Issue{x}
\Month{xx}
\Year{2012}
\pagespan{1}{}
\Receiveddate{Day Month 2012}
\Reviseddate{Day Month 2012}
\Accepteddate{Day Month 2012}
\Dateposted{Day Month 2012}



\title[Wrapping corrections beyond the $\mathfrak{sl}(2)$ sector in $\mathcal{N}=4$ SYM]{Wrapping corrections beyond the $\mathfrak{sl}(2)$ sector in $\mathcal{N}=4$ SYM}


\author[M. Beccaria]{Matteo Beccaria\inst{1,}%
  \footnote{\textsf{matteo.beccaria@le.infn.it}.
}}
\address[\inst{1}]{  Dipartimento di Fisica, Universita' del Salento, 
  Via Arnesano, 73100 Lecce \&\\
  INFN, Sezione di Lecce.
}
\author[G. Macorini]{Guido Macorini\inst{2,}
   \footnote{\textsf{macorini@nbi.ku.dk}.
}}
\address[\inst{2}]{Niels Bohr International Academy and Discovery Center,  \\
		Blegdamsvej 17 DK-2100 Copenhagen, Denmark.
}
\author[C. Ratti]{CarloAlberto Ratti\inst{1,}
  \footnote{\textsf{carloalberto.ratti@le.infn.it}.
}}
\begin{abstract}
The $\mathfrak{sl}(2)$ sector of $\mathcal{N} = 4$ SYM theory has been much studied and the anomalous dimensions of those operators are well known. 
Nevertheless, many interesting operators are not included in this sector. 
We consider a class of twist operators beyond the $\mathfrak{sl}(2)$ subsector introduced by Freyhult, Rej and Zieme. 
They are spin $n$, length-3 operators.
At one-loop they can be identified with three gluon operators.
At strong coupling, they are associated with spinning strings with two spins in $AdS$ space and charge in $S^5$. 
We exploit the Y-system to compute the leading weak-coupling four loop wrapping correction to their anomalous dimension. 
The result is written in closed form as a function of the spin $n$. 
We combine the wrapping correction with the known four loop  asymptotic Bethe Ansatz contribution and analyze special limits in the spin $n$. 
In particular, at large $n$, we prove that a generalized Gribov-Lipatov reciprocity holds.
At negative unphysical spin, we present a simple BFKL-like equation predicting the rightmost leading poles.
\end{abstract}
\maketitle                   

\section{Introduction}
The mirror thermodynamic Bethe Ansatz (TBA) for the $AdS_5\times S^5$ superstring \cite{TBA1} provides a general approach to the study of finite size corrections to states/operators in AdS/CFT.
The associated Y-system has been proposed in \cite{GKV} based on symmetry arguments and educated guesses about the analyticity and asymptotic properties of the Y-functions.

Very powerful explicit tests of these methods have been mostly done in the $\mathfrak{sl}(2)$ sector of $\mathcal{N}=4$ SYM \cite{TBA2}.
Beyond $\mathfrak{sl}(2)$, our knowledge of the larger part of the full $\mathfrak{psu}(2,2|4)$ structure of the theory is limited.

In this brief note, following \cite{noi1}, \cite{noi2}, we show some recent progress in the computation of the leading order wrapping corrections of the Freyhult-Rej-Zieme (FRZ) operators, a special class of operators beyond the $\mathfrak{sl}(2)$ sector first discussed at one-loop in the general analysis \cite{Niklas} and further analyzed at all orders in  \cite{FRZ}. 
The results follow from a direct application of the Y-system techniques.

FRZ operators are twist 3 operators. 
Their general form can be written by inserting covariant and anti-covariant derivatives $\mathscr{D}$, $\Bar{\mathscr{D}}$ into the half-BPS state $\mbox{Tr} \mathcal{Z}^{3}$ ($\mathcal Z$ being one of the three
complex scalars of $\mathcal N=4$ SYM theory):
\begin{equation}
  \label{operator}
  \mathbb{O}^{\rm FRZ}_{n,m} = \mbox{Tr}\,(\mathscr{D}^{n+m} \Bar{\mathscr{D}}^m \mathcal{Z}^3)+\cdots.
\end{equation}
At strong coupling, these operators are dual to spinning string configurations with two spins $S_1 = n+m-\frac{1}{2}$ and $S_2 = m-\frac{1}{2}$ in $AdS_{5}$ and charge $J=L=3$ in $S^5$.
The distribution of the roots on the nodes of the $\mathfrak{psu}(2,2|4)$ algebra is summarized in Figure \ref{eq:dynkin3}:
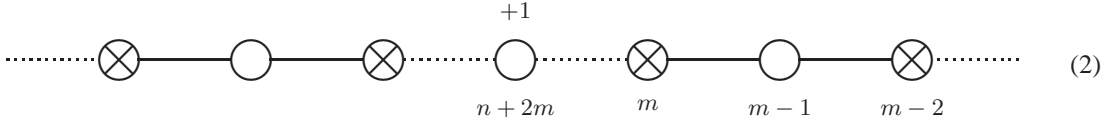
\begin{eqnarray}
  \label{eq:dynkin3}
  \phantom{a}
  \hspace{10mm}
  \begin{minipage}{260pt}
    \setlength{\unitlength}{1pt} \small\thicklines
    \begin{picture}(260,55)(-10,-30)
      \dottedline{3}(-72,0)(-40,0)  
      \put(-30,00){\circle{15}}
      \put(-35,-5){\line(1, 1){10}}  
      \put(-35, 5){\line(1,-1){10}}  
      \put(-23,00){\line(1,0){35}} 
      \put( 20,00){\circle{15}}     
      \put( 27,00){\line(1,0){35}} 
      \put( 70,00){\circle{15}}
      \put( 65,-5){\line(1, 1){10}}  
      \put( 65, 5){\line(1,-1){10}}  
      \dottedline{3}(78,0)(112,0) 
      \put(120,00){\circle{15}}
      \put(120,15){\makebox(0,0)[b]{$+1$}} 
      \put(120,-15){\makebox(0,0)[t]{$n+2m$}} 
      \dottedline{3}(128,0)(162,0) 
      \put(170,00){\circle{15}}
      \put(165,-5){\line(1, 1){10}}  
      \put(165, 5){\line(1,-1){10}}  
      \put(170,-15){\makebox(0,0)[t]{$m$}} 
      \put(177,00){\line(1,0){35}} 
      \put(220,00){\circle{15}}
      \put(220,-15){\makebox(0,0)[t]{$m-1$}} 
      \put(227,00){\line(1,0){35}} 
      \put(270,00){\circle{15}}
      \put(265,-5){\line(1, 1){10}} 
      \put(265, 5){\line(1,-1){10}} 
      \put(270,-15){\makebox(0,0)[t]{$m-2$}} 
      \dottedline{3}(310,0)(280,0) 
    \end{picture}
  \end{minipage}
\end{eqnarray}
Non trivial roots out of the $\mathfrak{sl}(2)$ sector occur for $m \geq 2$. In the particular $m=2$ case, the operators are called 3-gluon operators.

In general, the wrapping effects are expected to appear at weak coupling for the $\mathbb{O}^{\rm FRZ}_{n,m}$ operators at four loop.
In paper \cite{Baxter-poly}, the asymptotic part of the anomalous dimensions of the $\mathbb{O}^{\rm FRZ}_{n,2}$ has been computed exactly up to  four loops. 
The result is given as a closed formula in the spin parameter $n$. Up to this level, reciprocity holds for the asymptotic contributions.
By following \cite{noi1}, from the Y-system equations we derive precisely a similar closed formula for the leading wrapping corrections.
This formula, together with the results of \cite{Baxter-poly}, completes the study of the anomalous dimensions for the 3-gluon operators up four loop. 
As a byproduct, we shall be able to test positively reciprocity as well as discuss the BFKL poles of the full four loop result. 
We also show that a very simple and natural modification of the twist-2 BFKL equation predicts the correct pole structure.

For the general $\mathbb{O}^{\rm FRZ}_{n,m}$ FRZ operators, no asymptotic closed formula for the anomalous dimension is known beyond one-loop. 
Thus, the research of such a closed formula for the wrapping corrections is hopeless and useless.
Nevertheless the large $n$ expansion of the asymptotic minimal anomalous dimension of $\mathbb{O}^{\rm FRZ}_{n,m}$ is known for fixed ratio $n/m$ or fixed $m$ \cite{FRZ}. 
The expansion is obtained at all orders in the coupling and including the leading term $\sim \log n$ as well as the subleading asymptotically constant correction $\sim n^{0}$. 
These two contributions are expected to be free of wrapping corrections.

Following \cite{noi2}, we consider precisely the large $n$ expansion of leading order wrapping correction which appears at four loop. 
We provide an algorithm to compute through the Y-system the large $n$ expansion for fixed $m$ and present explicit new results for $m=2, 3, 4$. 
For $m=2$ we match the large $n$ expansion of the closed formula for the 3-gluon operators \cite{noi1}.
The expansions for the other two values are new. 
In full generality, we prove the $\frac{\log n}{n^{2}}$ scaling behaviour at large $n$ thus confirming the assumption in \cite{FRZ}.

The plan of the paper is the following: In the next Section we give the minimal set of Y-system equations required to compute the leading order wrapping corrections. 
In Section \ref{3glu} we give the result for the leading wrapping corrections of the 3-gluon operators.
In Section \ref{largen} we consider the large $n$ limit of  FRZ operators $\mathbb{O}^{\rm FRZ}_{n,m}$.
The general strategy is described and results are given for $m=(3,4)$.


\section{Leading wrapping corrections from Y-system}
\label{generalities}

The Y-system is a set of functional equations for the functions $Y_{a,s}(u)$ defined on the fat-hook of
$\mathfrak{psu}(2,2|4)$ \cite{GKV}.
These equations are
\begin{equation}
\label{Y-system}
\frac{Y^{+}_{a,s}\,Y^{-}_{a,s}}{Y_{a+1,s}\,Y_{a-1,s}}=\frac{
(1+Y_{a,s+1})(1+Y_{a,s-1})}{(1+Y_{a+1,s})(1+Y_{a-1,s})}.
\end{equation}
Their boundary conditions are discussed in \cite{Gromov:2009tv}.
The anomalous dimension of a generic state is given by the TBA formula 
\begin{equation}
  \label{eq:TBAenergy}
  E = \sum_{\ell = 0}^{\infty} g^{2 \ell} \gamma_{\ell\textrm{-loop}} =  \underbrace{\sum_{i} \epsilon_{1}(u_{4,i})}_{\rm asymptotic\ \gamma^{asy}}+\underbrace{\sum_{a\ge 1}\int_{\mathbb R}\frac{du}{2\pi i}
    \frac{\partial \epsilon_{a}^{\star}}{\partial u}\,\log(1+Y^{\star}_{a,0}(u))}_{\rm wrapping\ W},
\end{equation}
where the asymptotic and the wrapping contributions are well distinct. In formula (\ref{eq:TBAenergy}), the dispersion relation is 
\begin{equation}
\epsilon_{a}(u) = a+\frac{2\,i\,g}{x^{[a]}}-\frac{2\,i\,g}{x^{[-a]}},
\end{equation}
and the star means evaluation in the mirror kinematics\footnote{Shifted quantities are defined as 
$ F^{\underbrace{\pm\dots\pm}_{a}}(u) = F^{[\pm a]}(u) = F\left(u\pm i\,\frac{a}{2}\right).$}. 

The crucial assumption in the identification of relevant solutions to the Y-system is 
\begin{equation}
Y^{\star}_{a\ge 1, 0}\sim \left(\frac{x^{[-a]}}{x^{[+a]}}\right)^{L},
\end{equation}
for large $L$, or large $u$ (or small $g$). In this limit, it can be shown that the Hirota equation splits
in two $\mathfrak{su}(2|2)_{\rm L, R}$ wings. One can have a simultaneous finite large $L$ limit on both wings
after a suitable gauge transformation. The solution is then
\begin{equation}
\label{eq:Ysolution}
Y_{a,0}(u) \simeq  \left(\frac{x^{[-a]}}{x^{[+a]}}\right)^{L}\,\frac{\phi^{[-a]}}{\phi^{[+a]}}\,
T^{\rm L}_{a,1}\,T^{\rm R}_{a,1},
\end{equation}
where $\frac{\phi^{[-a]}}{\phi^{[+a]}}$ is the fusion form factor and $T^{\rm L, R}_{a, 1}$ are the transfer matrices of the antisymmetric rectangular representations of $\mathfrak{su}(2|2)_{\rm L, R}$.
They can be explicitly computed by an appropriate generating functional \cite{KSZ}.

At weak coupling, the formula (\ref{eq:TBAenergy}) for the leading wrapping corrections takes the simpler form
\begin{equation}
  \label{wrapping}
  W = -\frac{1}{\pi}\sum_{a=1}^{\infty}\int_{\mathbb R}du\,Y_{a,0}^{\star}
  = - 2 i \sum_{a=1}^{\infty} \textrm{Res}_{u=\frac{i a}{2}} Y_{a,0}^{\star}.
\end{equation}
The formula (\ref{eq:Ysolution}) for the relevant $Y$ functions gets simplified too.
In particular, the $Y$'s can be expressed as functions of the 1-loop Baxter polynomials $Q_i$'s. We get \cite{noi2}:
\begin{eqnarray}
  \label{dispersion}
  &&
   \left(\left(\frac{x^{[-a]}}{x^{[+a]}}\right)^{L}\right)^\star = \left(\frac{4 g^2}{ a^2 + 4 u^2}\right)^L,
   \nonumber \\
  &&
   \left(\frac{\phi^{[-a]}}{\phi^{[+a]}}\right)^\star
   = \left[Q_4^+(0)\right]^2 \frac{Q_4^{[1-a]}}{Q_4^{[-1-a]} Q_4^{[a-1]} Q_4^{[a+1]}} ~ 
  \frac{Q_5^{[-a]}}{Q_5(0)} ~
  \frac{Q_7(0)}{Q_7^{[-a]}},
  \nonumber \\
  &&  T^{*,L}_{a,1} =
  \left(\log\left(Q_4^\prime\right)\right)|^{u=+\frac{i}{2}}_{u=-\frac{i}{2}} 
  \ g^2 \frac{(-1)^{a+1}}{Q_4^{[1-a]}} 
  \mathop{\sum^{a}_{k=-a}}_{\Delta k =2}
  \frac{Q_4^{[-1-k]} - Q_4^{[1-k]}}{u-i\frac{k}{2}}\bigg|_{Q_4^{[-1-a]},Q_4^{[-1-a]}\rightarrow 0}  
  +\mathcal{O}(g^4),
  \nonumber \\
  &&
  T^*_{a,1} =
  (-1)^{a+1} \frac{Q_5^{[a]} Q_7^{[-a]}}{Q_4^{[1-a]}}  
  \mathop{\sum^{a-1}_{k=1-a}}_{\Delta k=2}
   \frac{Q_4^{[k]}}{Q_6^{[k]}} \left(\frac{Q_6^{[k+2]}-Q_6^{[k]}}{Q_5^{[k+1]} Q_7^{[k+1]}}
    + \frac{Q_6^{[k-2]}-Q_6^{[k]}} {Q_5^{[k-1]} Q_7^{[k-1]}}\right) +\mathcal{O}(g^2).
\label{formulas}
\end{eqnarray}
The expressions for the $Q_i$'s for the FRZ operators are known and can be found in \cite{FRZ}, \cite{Baxter-poly}. 
These expressions together with equations (\ref{eq:Ysolution}), (\ref{wrapping}) and (\ref{formulas}) are in principle all what one needs to compute the leading wrapping corrections for the FRZ operators.

\section{The 3-gluon operators $\mathbb{O}^{\rm FRZ}_{n,2}$ case}
\label{3glu}
FRZ operators $\mathbb{O}^{\rm FRZ}_{n,2}$  with $m=2$ are among the simplest non trivial operators that can be built beyond the $\mathfrak{sl}(2)$ sector of $\mathcal{N}=4$ SYM theory.
In this Section we give a general result for their leading order wrapping corrections $W_n$. 
More details about the results of this section can be found in \cite{noi1}.

The starting point is the observation that by specializing the formulas of Section \ref{generalities} to the simple $m=2$ case and fixing $n$, the computation of the wrapping corrections is straightforward. 
In \cite{noi2} we produced a list of results for the wrapping up to $n=70$. Then we were able to condensate these data in a closed formula that replicates and generalizes to any value of $n$ the numerical results of the list.
The formula is
\begin{eqnarray}
\label{formula}
W_{n} &=& \left(r_{0,n}+r_{3,n}\,\zeta_3+r_{5,n}\,\zeta_5\right)\,g^{8},
\nonumber \\
r_{5,n} &=& 80\, \left(4\,S_{1}+\frac{2}{N+1}+4\right)\,\left(-4(N+1)+\frac{1}{N+1}\right),
\nonumber \\
r_{3,n} &=& 16\, \left(4\,S_{1}+\frac{2}{N+1}+4\right)\times 
\nonumber \\
&& \quad \ \,\left[
8\,(N+1)\,S_{2}+8+ \frac{2}{N+1}(2-S_{2}) -\frac{2}{(N+1)^{2}}-\frac{1}{(N+1)^{3}}
\right],
\nonumber\\
r_{0,n} &=& 2\, \left(4\,S_{1}+\frac{2}{N+1}+4\right)\times  \nonumber \\
&& \quad \ \,\left[ 16\,(N+1)\,(2\,S_{2,3}-S_{5})+32\,S_{3}+\frac{4}{N+1}(S_{5}-2S_{2,3}+4S_{3})+ \right.
\nonumber\\
&& \quad \  \left. +\frac{8}{(N+1)^{2}}(-S_{3}+2)+\frac{4}{(N+1)^{3}}(-S_{3}+4)-\frac{4}{(N+1)^{5}}-\frac{1}{(N+1)^{6}}
\right].
\end{eqnarray}
Here $S_{a,b,\dots}\equiv S_{a,b,\dots}(N)$ and $N = n/2 +1$\footnote{We remind that $n$ is an even positive integer. This means that $N$ is an integer, $N\geq 2$.}. 
Note that each of the rational coefficient $r_i$ can be written as $r_{i,n}=\gamma_{\textrm{1-loop}} \widetilde{r}_{i,n}$, 
where $\gamma_{\textrm{1-loop}}= S_{1}+ \frac{2}{N+1}+4$ is the one loop anomalous dimension.

The Ansatz (\ref{formula}) completes the four loop expression of the energy spectrum for the $\mathbb{O}^{\rm FRZ}_{n,2}$ operators, 
the other relevant contributions up to this order being the first four asymptotic orders. Their expressions can be found in \cite{Baxter-poly}.

Up to four loop, the asymptotic part of the spectrum shows the generalized Gribov-Lipatov reciprocity property \cite{Baxter-poly}.
Formula (\ref{formula}) allows to check whether this property extends to the full four loop result. 
This is indeed the case: The large $n$ expansion of $W_n/\gamma_{\textrm{1-loop}}$ reads
\begin{eqnarray}
\lefteqn{\zeta_{5}\,\widetilde r_{5, n} +\zeta_{3}\,\widetilde r_{3, n}+\widetilde r_{0, n} = } && 
\nonumber \\
&& \frac{\frac{32}{3}-32 \zeta _3}{J^2}+\frac{\frac{232 \zeta _3}{5}-\frac{352}{15}}{J^4}+\frac{\frac{4834}{105}-\frac{2344 \zeta _3}{35}}{J^6}+\frac{\frac{3544
   \zeta _3}{35}-\frac{83956}{945}}{J^8}+\frac{\frac{271768}{1485}-\frac{9512 \zeta _3}{55}}{J^{10}}+
   \nonumber \\
   && \frac{\frac{1872392 \zeta
   _3}{5005}-\frac{20053258}{45045}}{J^{12}}+\frac{\frac{87933002}{61425}-\frac{524872 \zeta _3}{455}}{J^{14}}+\frac{\frac{4917304 \zeta
   _3}{935}-\frac{5747755528}{883575}}{J^{16}}+\dots, 
\end{eqnarray}
where we introduced the charge $J^2 = N (N+2)$. All the odd powers of $1/J$ cancel proving that the reciprocity property does hold. 
It is remarkable that this property is a consequence of non-trivial cancellations of odd $1/J$ terms which are present in the expansion of each single coefficient $\widetilde{r}_{i,n}$.
The presence of reciprocity is really appreciable, since it allows to predict a half of the large $n$ expansion terms (expressed as functions of the same $n$) as combinations of the other half.

The Ansatz (\ref{formula}) allows also to study the BFKL poles of the full four loop result. 
In general, at $\ell$-loop the analytic continuation of $\gamma_{\ell\textrm{-loop}}$ in the variable $N$ around $N=-1$ is expected to behave at worst as $\omega^{-\ell}$, where $\omega$ is a small expansion parameter defined by $N=-1+\omega$.
At four loops the asymptotic anomalous dimension $\gamma_{\textrm{4-loop}}^{asy}$ presents instead also poles in $\omega^{-k}$ with $k=(7,6,5)$.
These poles get indeed compensated inside the full $\gamma_{\textrm{4-loop}}=\gamma_{\textrm{4-loop}}^{asy}+W$ where precisely the wrapping contribution (\ref{formula}) is included.
The final expressions for the expansions of the first four $\gamma_{\ell\textrm{-loop}}$  are
\begin{eqnarray}
&& \gamma_{\textrm{1-loop}} = -\frac{4}{\omega }+\dots, 
\qquad \qquad \qquad \qquad \quad \ \ \ \quad
\gamma_{\textrm{2-loop}} = \frac{8}{\omega ^2}+\frac{4 \pi ^2}{3 \omega }+\dots, 
\\
&& \gamma_{\textrm{3-loop}} = \frac{\bf 0}{\omega^{3}}-\frac{16 \left(-3 \zeta _3+\pi ^2+12\right)}{3 \omega ^2}+\dots,
\quad 
\gamma_{\textrm{4-loop}} = -\frac{32 \left(1 + 2 \zeta _3\right)}{\omega ^4}+\frac{160 \zeta _3}{\omega ^3}+\dots.
\nonumber
\end{eqnarray}
Strikingly, the leading poles  can be reproduced by a BFKL-like equation that links $\omega$ to the full anomalous dimension $\gamma$
\begin{eqnarray}
\label{bfkl}
&&
-\frac{\omega}{g^{2}} = \,\chi_{1}\left(\frac{\gamma}{2}\right), \qquad \qquad \chi_{1}\left(z\right) = S_{1}(z) + S_{1}(z+1).
\end{eqnarray}
In fact, the weak coupling expansion of this equation reads precisely
\begin{eqnarray}
\gamma &=& \left(-\frac{4}{\omega}+\dots\right)\,g^{2}+
\left(\frac{8}{\omega^{2}}+\dots\right)\,g^{4}+
\left(\frac{\mathbf{0}}{\omega^{3}}+\dots\right)\,g^{6}+
\left(-\frac{32\,(1+2\,\zeta_{3})}{\omega^{4}}+\dots\right)\,g^{8}
+ \dots.
\nonumber \\
\end{eqnarray}

\section{General FRZ operators: Large $n$ expansion}
\label{largen}
For the general $\mathbb{O}^{\rm FRZ}_{n,m>2}$ only the large $n$ limit of the asymptotic part of the anomalous dimension is known for fixed ratio $n/m$ or fixed $m$ \cite{FRZ}.
Thus, as far as we consider the wrapping corrections, we are primarily interested in the computation of their large $n$ expansions.

The Y-system described in Sec. \ref{generalities} can be optimized to get only the large $n$ contributions. The computational strategy is the following:
\begin{itemize}
\item Starting from eq. (\ref{wrapping}) one evaluates the residue at fixed $a=1,
2, \dots$ without assigning $n$;
\item Analyzing the dependence on $n$ of the residue, one realizes that $n$ comes from the Baxter polynomials
$Q_{4}$, its derivatives (which are written in terms of the basic hypergeometric function $F_{n,m}$) and
from the explicit $n$-dependent coefficients of the other Baxter polynomials. 
At this point the limit over $n$ can be taken in two distinct steps;
\item Using the Baxter equation, it is possible to shift
the argument of $F_{n,m}$ to some minimal value and take the large $n$
limit on the coefficients. This gives a first expansion containing
various derivatives of the logarithm of $F_{n,m}$;
\item The derivatives of the logarithm can be
systematically computed by means of the method explained in \cite{Beccaria:2007uj};
\item The outcome of this procedure are sequences of rational numbers being the
$a$-dependent coefficients of the large $n$ expansion of $\mathop{\textrm{Res}}_{u=i\frac{a}{2}} Y_{a,0}^{\star}$. 
These sequences turn out to be rather simple rational functions which are easily identified. 
The sum over $a$ of these rational functions gives the wrapping $W$ as defined by eq. ($\ref{wrapping}$).
\end{itemize}

Following this computational method, in \cite{noi2} we computed  the large $n$ expansion of the wrapping corrections for the operators $\mathbb{O}^{\rm FRZ}_{n,m}$ with $m=2,3,4$.
The $m=2$ case agrees with the large $n$ expansion of eq. (\ref{formula}). For $m=3,4$ we get these new results
\begin{eqnarray}
  g^{-8}\,W_{n,m=3} &=& -\frac{4}{3}\, (36\,\zeta_{3}+5\pi^{2}-3)\,\frac{2\,\log{\bar n}+1}{n^{2}}
  +\frac{88}{3}\, (36\,\zeta_{3}+5\pi^{2}-3)\,\frac{\log\bar n}{n^{3}}+\\
  && -\frac{2}{9\,n^{4}}\left[
    4\,(9108\,\zeta_{3}+1289\pi^{2}-615)\,\log\bar n-14868\,\zeta_{3}-2017\pi^{2}+1527
  \right]+\dots, \nonumber\\
  g^{-8}\,W_{n,m=4} &=& -\frac{1024}{1215}\, (81\,\zeta_{3}-32)\,\frac{3\,\log{\frac{\bar n}{2}}+4}{n^{2}}
  +\frac{7168}{1215}\, (81\,\zeta_{3}-32)\,\frac{6\,\log\frac{\bar n}{2}+5}{n^{3}}+\\
  && -\frac{1024}{8505\,n^{4}}\left[
    138240\,(1971\,\zeta_{3}-760)\,\log\frac{\bar n}{2}+143289\,\zeta_{3}-53248\right]+\dots. \nonumber
\end{eqnarray}
The extension of these results to larger values of $m$ is a plain task.

The numerical check in Table (\ref{tab1}) shows that there is good agreement between our expansions and a numerical estimate of the wrapping corrections.
\begin{equation}
\label{tab1}
  \begin{array}{|c||c|c|ccc|c|c|}
  \hline
  m=3&  n & \mbox{estimate} & ({\rm LO} & {\rm NLO} & {\rm NNLO}) & \mbox{full expansion} & \mbox{diff \%}\\
    \hline
   & 10 & -1.857411 & (-4.038730 & 3.785374 & - 2.548066) & -2.801422  & 51 \%\\
   & 30 & -0.438781 & (-0.594614 & 0.193683 & -0.045355) &  -0.446286  & 1.7 \%\\
   & 50 & -0.197270 & (-0.238478 & 0.047207 & - 0.006716)  & -0.197986 & 0.36 \%\\
\hline
  m=4&  n & \mbox{estimate} & ({\rm LO} & {\rm NLO} & {\rm NNLO}) & \mbox{full expansion} & \mbox{diff \%}\\
    \hline
    & 10 & -1.201460 & ( - 2.908787 & 3.493848 &  - 3.576129 ) &  -2.991068 & 149 \% \\
    & 30 & -0.293765 & (- 0.424071 & 0.176476 & - 0.061888) &  -0.309484 & 5.4\% \\
    & 50 & -0.134246 & (- 0.169551 & 0.042847 & - 0.009090 ) & - 0.135794 & 1.2 \%\\
\hline
  \end{array}
\end{equation}

Following the general idea of reciprocity, we are led to rewrite the above large $n$ expansions in terms of the quantity
\begin{equation}
  J^{2}_{m} = n\,(n+a_{m}).
\end{equation}
It turns out that the coefficients of the two odd terms $1/ J^{3}$ and
$\log J/ J^{3}$ indeed vanish for the choice $ a_{2,
  3, 4} = 8, \ 11, \ 14$. This is not completely trivial since we
have one parameter and two structures.  It is tempting to conjecture
the simple relation $a_{m} = 3\,m+2$ and to claim that reciprocity in
the above sense holds for the full anomalous dimension as well.


\end{document}